\documentclass[12p]{article}
\usepackage{latexsym}
\usepackage{color}
\usepackage{amsmath}
\usepackage{epsfig}
\usepackage{hyperref}
 \usepackage{dcolumn}
   \usepackage{threeparttable}
     \newcommand{\thickhline}{\noalign{\hrule height 0.8pt}}
\newcommand{\ortala}[1]{\begin{center}#1\end{center}}

\newcommand{\sandd}[1]{\left\langle #1\right\rangle}

\newcommand{\summ}[3]{{{\underset{#1 }{\overset{#2}{\displaystyle\sum}}}#3}}

\newcommand{\re}[1]{(\ref{#1})}

\newcommand{\eq}[2]{\begin{equation}\label{#1}  #2\end{equation}}

\newcommand{\paran}[1]{\left(#1\right)}

\newcommand{\sch}[1]{Schrodinger}

\newcommand{\komb}[2]{\paran{\begin{array}{c} #1 \\ #2 \end{array}}}

\linethickness{0.6pt}

\begin{document}

\ortala{\textbf{Hysteresis behavior of the anisotropic quantum Heisenberg model}}

\ortala{\textbf{\"Umit Ak\i nc\i \footnote{umit.akinci@deu.edu.tr}}}

\ortala{\textit{Department of Physics, Dokuz Eyl\"ul University,
TR-35160 Izmir, Turkey}}

\section{Abstract}

The effect of the anisotropy in the exchange interaction on the hysteresis loops within the anisotropic quantum Heisenberg model has been
investigated with the effective field theory for two spin cluster. Particular attention has been devoted on the behavior of the hysteresis loop area, coercive field and remanent magnetization with the anisotropy in the exchange interaction for both ferromagnetic and paramagnetic phases.

Keywords: \textbf{Quantum anisotropic Heisenberg model; hysteresis loops; hysteresis loop area; coercive field; remanent magnetization}

\section{Introduction}\label{introduction}

Hysteresis is a common behavior of the most of the physical systems. It originates from the delay of the response of the system to the driving cyclic force.  This fact also shows itself in the magnetic systems in the presence of varying magnetic field. Magnetic hysteresis is one of the most important  features of the magnetic materials. The shape of the hysteresis loops can be determined by three properties;  namely, hysteresis loop area (HLA), coercive field (CF) and remanent magnetization (RM). It is well known fact that these  hysteresis properties are very important in manufacturing of magnetic recording media. The RM is defined as residual magnetization which is the magnetization
left behind in the system after an external magnetic field is removed and CF is defined as the intensity of the external magnetic field needed to change the sign of the magnetization. On the other hand,  HLA corresponds to energy loss due to the hysteresis.

When we compare with the Ising Model, as far as we know, there has been less attention paid
on the hysteresis behavior in the Heisenberg model which is a more realistic
model than the Ising model for the spin systems. Indeed, anisotropic or isotropic Heisenberg antiferromagnet in magnetic
field is widely studied since magnetic field in antiferromagnetic systems cannot destroy the phase transition, conversely can create different phases. For instance, one dimensional anisotropic spin-1/2 Heisenberg model in transverse and longitudinal magnetic field   \cite{ref1,ref2} and two dimensional spin-1/2 anisotropic antiferromagnetic
Heisenberg model in magnetic field \cite{ref3,ref4} have been studied. In a ferromagnetic case, the magnetic properties of two dimensional mixed spin anisotropic ferromagnetic
Heisenberg model in a transverse magnetic field \cite{ref5} and longitudinal magnetic field \cite{ref6} have been investigated.
Also, zero-temperature hysteresis in random-field XY and Heisenberg models have been examined\cite{ref7,ref8}.
On the other hand, there are several works related to the kinetic Heisenberg model, e.g. dynamical hysteresis and scaling of the kinetic Heisenberg model \cite{ref9,ref10,ref11}.

Although there are works related to the  Heisenberg model in a magnetic field, as far as we know, less attention paid on the hysteresis characteristics of anisotropic Heisenberg model. Thus, the effect of the anisotropy in the exchange interaction on the hysteresis characteristics of the quantum Heisenberg model is the topic of this work. In this work we use the effective field theory (EFT) for the two spin cluster, which is abbreviated as EFT-2 in the literature.
EFT approximation can provide results that are superior to those obtained
within the traditional mean field approximation due to the taking into account the self spin correlations which are omitted in the mean field approximation.
One of the useful formulation within the EFT, namely differential operator technique is introduced by Honmura and Kaneyoshi for Ising systems \cite{ref12}. EFT for the typical Ising system starts from the constructing finite cluster of spins which represents the system. Callen-Suzuki spin identities \cite{ref13,ref14} are the starting point of the EFT for the one spin clusters. When one expands these identities with differential operator technique, multi spin correlations appear and in order to avoid from the mathematical difficulties, these multi spin correlations often neglected by using decoupling approximation \cite{ref15}. Working with larger finite clusters will give more accurate results. The  Callen-Suzuki identities have been generalized to two spin clusters in Ref. \cite{ref16} (EFT-2 formulation). This EFT-2 formulation has been successfully applied to the variety of systems such as quantum spin-1/2 Heisenberg ferromagnet \cite{ref17,ref18}  and antiferromagnet \cite{ref19},  classical n-vector model \cite{ref20,ref21}, as well as spin-1 Heisenberg ferromagnet \cite{ref22,ref23}. We follow the EFT-2 formulation which is derived in Ref. \cite{ref17} for this system.

The paper is organized as follows: In Sec. \ref{formulation} we
briefly present the model and  formulation. The results and
discussions are presented in Sec. \ref{results}, and finally Sec.
\ref{conclusion} contains our conclusions.

\section{Model and Formulation}\label{formulation}

We consider a lattice which consist of $N$ identical spins (spin-$1/2$) such that each of the spins has $z$ nearest neighbors. The Hamiltonian of the system is given by
\eq{denk1}{\mathcal{H}=-\summ{<i,j>}{}{\paran{J_x s_i^xs_j^x+J_y s_i^ys_j^y+J_z s_i^zs_j^z}}-\summ{i}{}{H_is_i^z}}
where $s_i^x,s_i^y$ and  $s_i^z$ denote the Pauli spin operators at a site $i$, $J_x,J_y$ and $J_z$ stand for the anisotropy in the exchange interactions between the nearest neighbor spins and $H_i$ is the longitudinal magnetic field at a site $i$. The first sum is over the nearest neighbors of the lattice, while the second one is over all the lattice sites.

We use two spin cluster approximation as an EFT formulation \cite{ref17}. In this approximation, we choose two spins (namely $s_1$ and $s_2$) and treat interactions exactly in this two spin cluster. In order to avoid some mathematical difficulties we replace the perimeter spins of the two spin cluster by Ising spins (axial approximation) \cite{ref17}. After all with using the differential operator technique and decoupling approximation (DA) \cite{ref15},
we obtain the magnetization per spin expression as
\eq{denk2}{
m=\sandd{\frac{1}{2}\paran{s_1^z+s_2^z}}=\sandd{\left[A_{x}+m B_{x}\right]^{z_0}
\left[A_{y}+m B_{y}\right]^{z_0}
\left[A_{xy}+m B_{xy}\right]^{z_1}} f\paran{x,y,H_1,H_2}|_{x=0,y=0}
} where each of $s_1$ and $s_2$ has number of $z_0$ distinct nearest neighbors and both of them have $z_1$ common nearest neighbor.

The coefficients are defined by
\eq{denk3}{
\begin{array}{lcl}
A_{x}=\cosh{\paran{J_z\nabla_x}}&\quad&
B_{x}=\sinh{\paran{J_z\nabla_x}}\\
A_{y}=\cosh{\paran{J_z\nabla_y}}
&\quad&
B_{y}=\sinh{\paran{J_z\nabla_y}}\\
A_{xy}=\cosh{\left[J_z\paran{\nabla_x+\nabla_y}\right]}&\quad&
B_{xy}=\sinh{\left[J_z\paran{\nabla_x+\nabla_y}\right]}\\
\end{array}
}
where $\nabla_x=\partial/\partial x$ and $\nabla_y=\partial/\partial y$ are the usual differential operators in the
differential operator technique. Differential operators act on an arbitrary function $g$ via
\eq{denk4}{\exp{\paran{a\nabla_x+b\nabla_y}}g\paran{x,y}=g\paran{x+a,y+b}}
for arbitrary constants  $a$ and $b$.

The function in Eq. \re{denk2} is given by
\eq{denk5}{f\paran{x,y,H_1,H_2}=\frac{x+y+H_1+H_2}{X_0}\frac{\sinh{\paran{\beta X_0}}}{\cosh{\paran{\beta X_0}}+\exp{\paran{-2\beta J_z}}\cosh{\paran{\beta Y_0}}}} and
\eq{denk6}{
X_0=\left[\paran{J_x-J_y}^2+(x+y+H_1+H_2)^2\right]^{1/2}, \quad Y_0=\left[\paran{J_x+J_y}^2+(x-y+H_1-H_2)^2\right]^{1/2}
}
where $\beta=1/(k_B T)$, $k_B$ is the Boltzmann
constant and $T$ is the temperature.

With the help of the Binomial expansion, Eq. \re{denk2} can be written as
\eq{denk7}{
m=\summ{p=0}{z_0}{}\summ{q=0}{z_0}{}\summ{r=0}{z_1}{}C^\prime_{pqr}m^{p+q+r}
} where the coefficients are
\eq{denk8}{
C^\prime_{pqr}=\komb{z_0}{p}\komb{z_0}{q}\komb{z_1}{r}A_x^{z_0-p}A_y^{z_0-q}A_{xy}^{z_1-r}B_x^{p}B_y^{q}B_{xy}^{r}f\paran{x,y,H_1,H_2}|_{x=0,y=0}
} and these coefficients can be calculated by using the definitions given in Eqs. \re{denk3} and \re{denk4}. Eq. \re{denk7} is often written in a more familiar form as
\eq{denk9}{
m=\summ{k=0}{z}{}C_{k}m^{k}
} and
\eq{denk10}{
C_{k}=\summ{p=0}{z_0}{}\summ{q=0}{z_0}{}\summ{r=0}{z_1}{}\delta_{p+q+r,k}C^\prime_{pqr}
} where $\delta_{i,j}$ is the Kronecker delta.

If we assume that the system is under the influence of an homogenous magnetic field (i.e. $H_1=H_2=H$ in Eq. \re{denk5}), for a given set of Hamiltonian parameters ($J_x,J_y,J_z$), as well as temperature and magnetic field ($H$), by determining the coefficients  from Eq. \re{denk8} we can obtain
a non linear equation from Eq. \re{denk9}, and by solving this equation we can get the magnetization ($m$) for given parameters. Then  we can construct the hysteresis loops which are nothing but the variation of the $m$ with $H$, by sweeping the magnetic field value from $-H$ to $H$ and then from $H$ to $-H$ again, and by calculating the magnetization at each magnetic field value with the procedure explained above. Hereafter,  once the hysteresis loop curve is determined, HLA, RM and CF can be obtained numerically.

\section{Results and Discussion}\label{results}
Let us the scale exchange interaction components with the unit of energy $J$ as,
\eq{denk11}{
r_n=\frac{J_n}{J}
}where $n=x,y,z$. Let us choose $r_z=1$, then $r_x,r_y$ can be used as the measure of the anisotropy in the exchange interaction. It can be seen from the definition of the function given in Eqs. \re{denk5} and  \re{denk6} that the transformation $J_x\rightarrow J_y,J_y\rightarrow J_x$ does not change the function. Since the factors $J_x,J_y$ enters the formulation only via this function we can say that mentioned transformation does not change the formulation. Hence, let us fix $r_x$  and concentrate only on varying $r_y$. Our investigation will be for honeycomb ($z_0=2,z_1=0$), Kagome ($z_0=2,z_1=1$) and  square ($z_0=3,z_1=0$) lattices.


\subsection{Isotropic Model}

In this case all components of the exchange interaction are equal to each other, i.e. $r_n=1$ in Eq. \re{denk11}. In order to represent the  behavior of the hysteresis loops with temperature for the isotropic quantum Heisenberg model, we depict the hysteresis loops for three different lattices with selected values of the temperature. The critical temperatures ($T_c$) of the these lattices can be seen in Table \ref{table1}.

\begin{table}[h]\label{table1}
\begin{center}
\begin{threeparttable}
\caption{The critical temperatures of the isotropic quantum Heisenberg model in the absence of magnetic field.}
\renewcommand{\arraystretch}{1.3}
\begin{tabular}{lllllllll}
\thickhline
Lattice & $z_0$& $z_1$ &$T_c$\cite{ref17}\\
\hline
Honeycomb & 2& 0 & 1.718 \\
Kagome & 2& 1 & 2.718 \\
Square& 3& 0 & 2.815 \\
\thickhline \\
\end{tabular}
\end{threeparttable}
\end{center}
\end{table}

\begin{figure}[h]\begin{center}
\epsfig{file=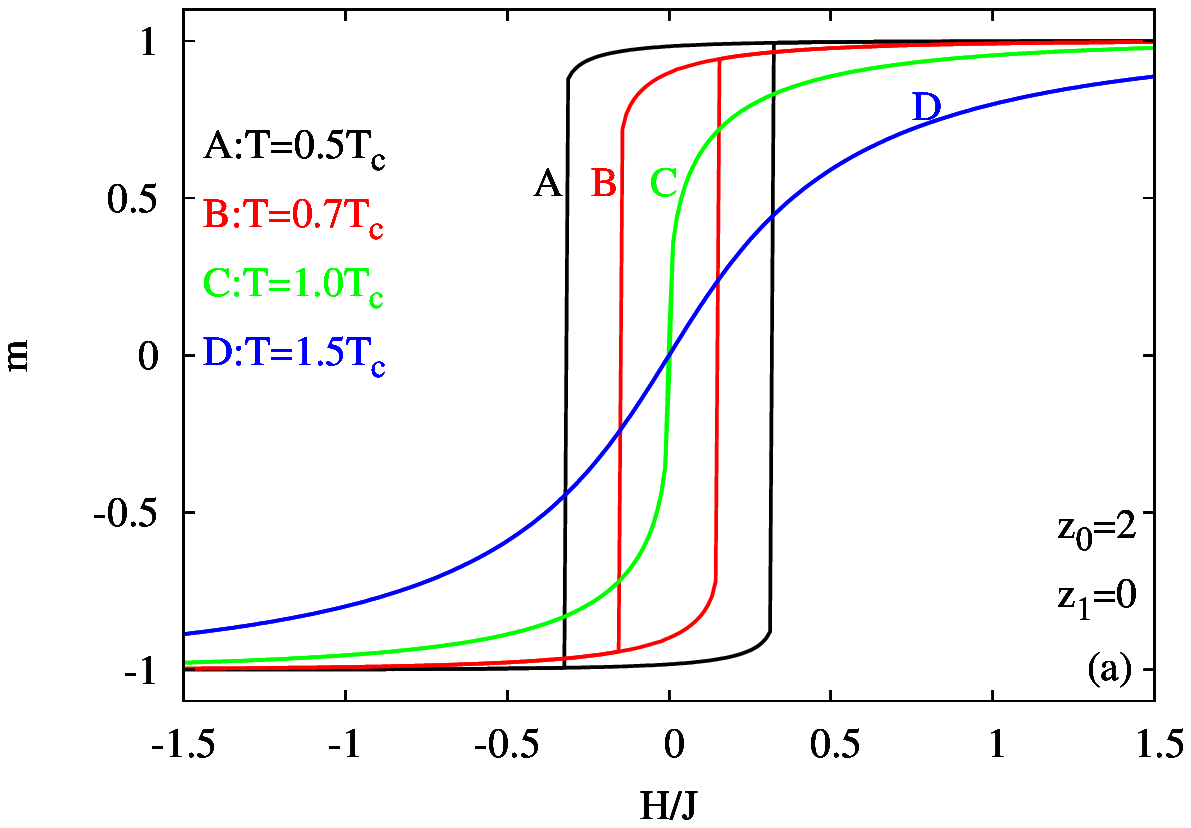, width=7.0cm}
\epsfig{file=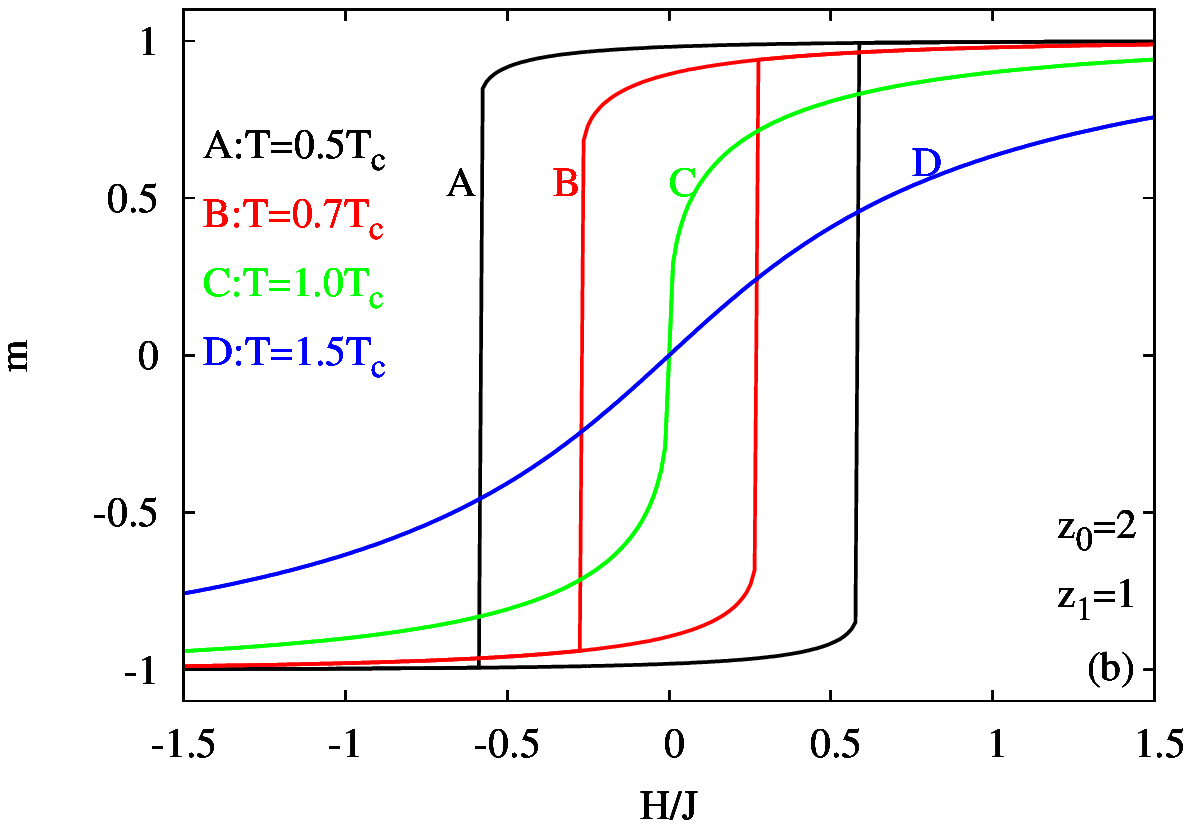, width=7.0cm}
\epsfig{file=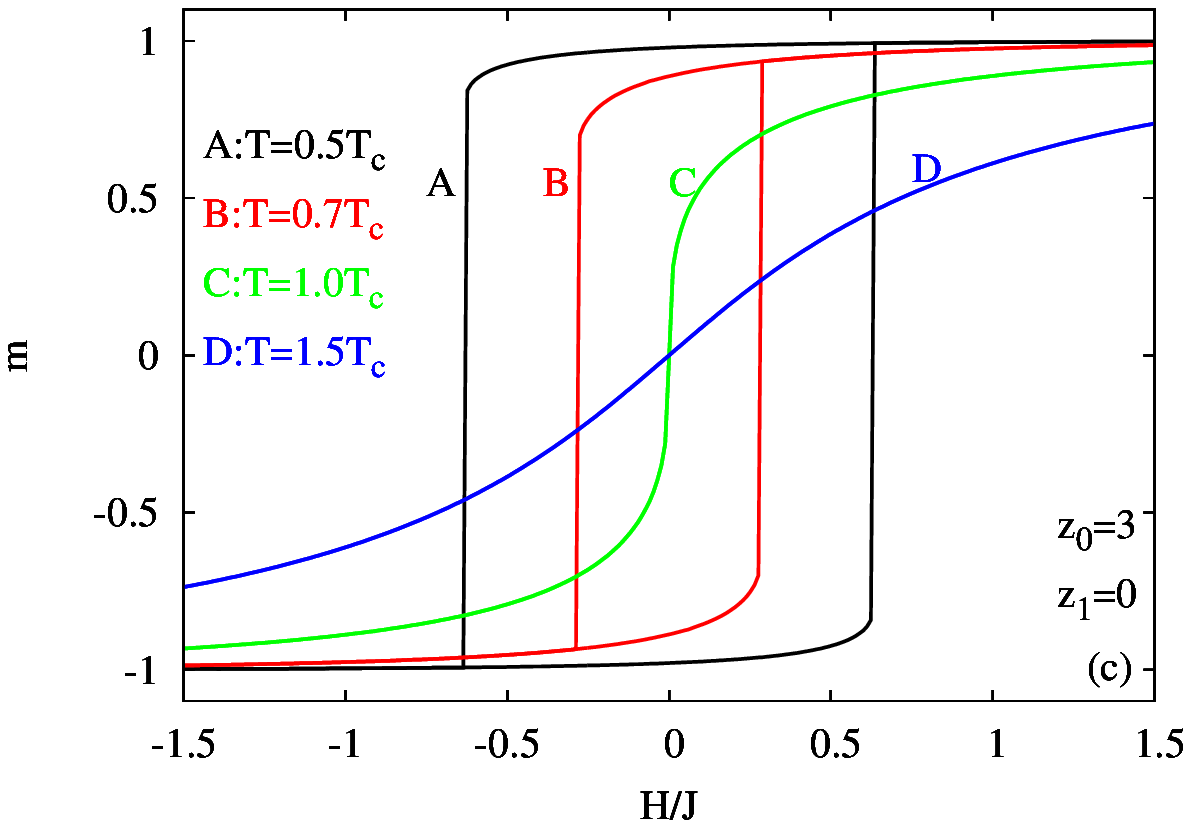, width=7.0cm}
\end{center}
\caption{Hysteresis loops for the isotrpopic quantum Heisenberg model with selected values of temperature for (a) honeycomb (b) Kagome and (c) square lattices.} \label{sek1}\end{figure}

As seen in Fig. \ref{sek1}, hysteresis loops corresponding to three lattices have finite HLA, CF  and RM in the ferromagnetic phase (i.e. $T<T_c$). As the temperature increases then the hysteresis loops evolve from loops labeled A to D in each figure. One can see from Fig. \ref{sek1} that, at a certain reduced temperature ($T/T_c$), square lattice has a wider HLA than that of the Kagome lattice and  Kagome lattice has a wider HLA than that of the honeycomb lattice (e.g. compare hysteresis loops labeled by B in Figs. \ref{sek1} (a),(b) and (c)) in the ferromagnetic phase. The same quantitative relationship is also valid for the CF and RM as seen in Fig. \ref{sek1}. This means that maximum energy loss due to the hysteresis occurs for a square lattice at a certain reduced temperature in the ferromagnetic region. The same relation is also expected for the CF curves concerning three different lattices at the same reduced temperature, since the 'resistivity' against the magnetic field of the square lattice, which comes from the spin-spin interaction, is stronger than that of the other two lattices.

As the temperature increases, when the system passes to a  paramagnetic phase from the ferromagnetic phase, the hysteresis behaviors of these lattices disappear. At a certain value of the magnetic field (which is different from zero) the response (i.e. magnetization) of the square lattice to the magnetic field is weaker than the Kagome lattice due to the strength of the spin-spin interactions. The same relation is also valid for the Kagome and honeycomb lattices. If we fit the $m-H$ curves in the paramagnetic region to the $\tanh{\paran{aH}}$ -where $a$ is the fitting parameter- we can interpret the harmony of the magnetization with the magnetic field i.e. magnitude of response of the system to the magnetic field. In this context, the greater value of $a$ means that, $m-H$ curve is similar to the Heaviside function. In this case magnitude of the response of the system to the magnetic field is high, and the magnetization is in harmony with the magnetic field. On the other hand, in the case of smaller value of $a$, the system cannot respond to the magnetic field due to the great thermal fluctuations.

\begin{figure}[h]\begin{center}
\epsfig{file=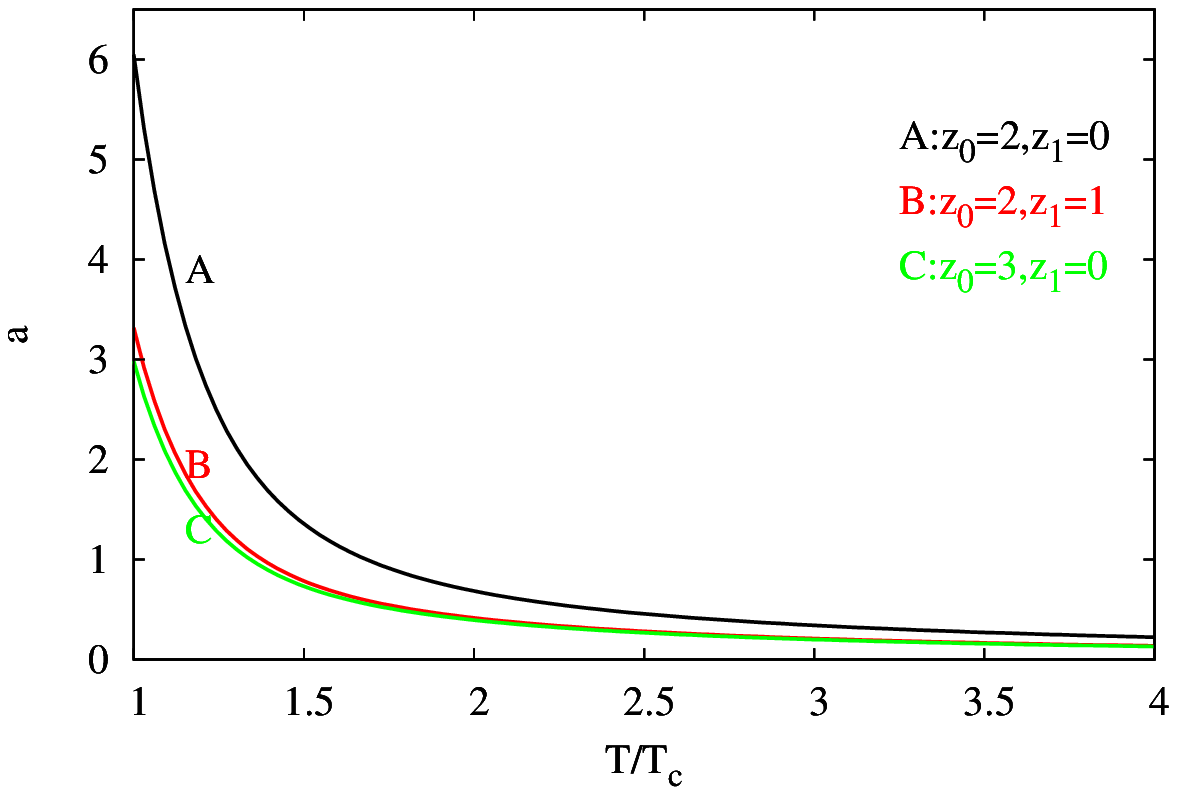, width=7.0cm}
\end{center}
\caption{Variation of the $\tanh{(aH)}$ fit results of the hysteresis loops for the isotropic quantum Heisenberg model in paramagnetic phase with the reduced temperature ($T/T_c$) for the (A) honeycomb (B) Kagome and (C) square lattices.} \label{sek2}\end{figure}

In Fig. \ref{sek2} we can see the variation of the parameter $a$ with the reduced temperature for three different lattices, in paramagnetic phase. At higher values of the reduced temperature the value of $a$  becomes close to zero which means that the curves become parallel lines with the $H$ axes, i.e. changing magnetic field makes no significant effect on the magnetization of the system due to the large thermal energy, which dictate the system to be in a disordered phase. Beside this, we can see from Fig. \ref{sek2} that, from relation of the magnitudes of $a$ at a certain reduced temperature, honeycomb lattice is the most affected lattice from the magnetic field, as expected.

\subsection{Anisotropic Model}

In this section, we want to investigate the effect of the anisotropy in the exchange interaction on the hysteresis loops in ferromagnetic phase. In order to achieve this purpose, let us choose the value of the reduced temperature as $T/T_c=0.5$ (where $T_c$ is the critical temperature of the related isotropic model with no magnetic field) and investigate the variation of the HLA, CF and RM with the anisotropy in exchange interaction. Since CF and RM adjust the width and the height of the hysteresis loop respectively, it is sufficient to investigate these properties with the HLA instead of drawing hysteresis loops.

\begin{figure}[h]\begin{center}
\epsfig{file=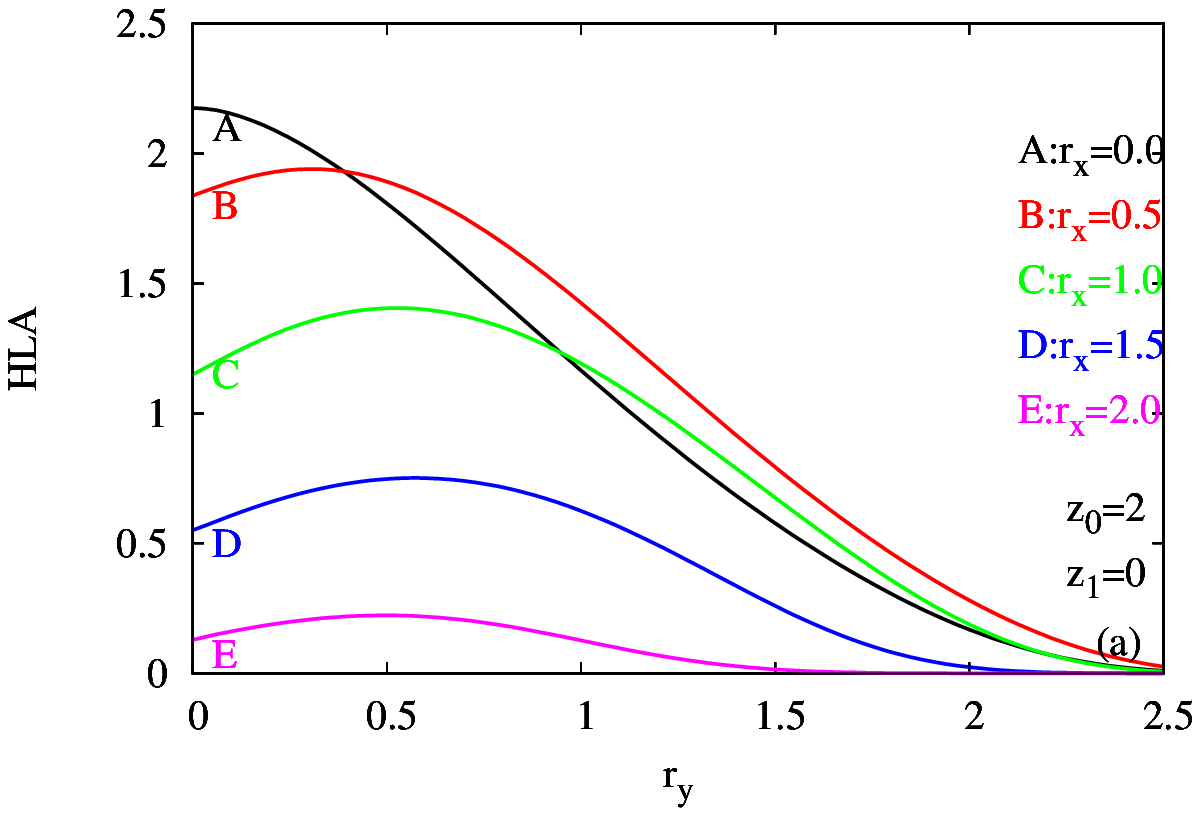, width=7.0cm}
\epsfig{file=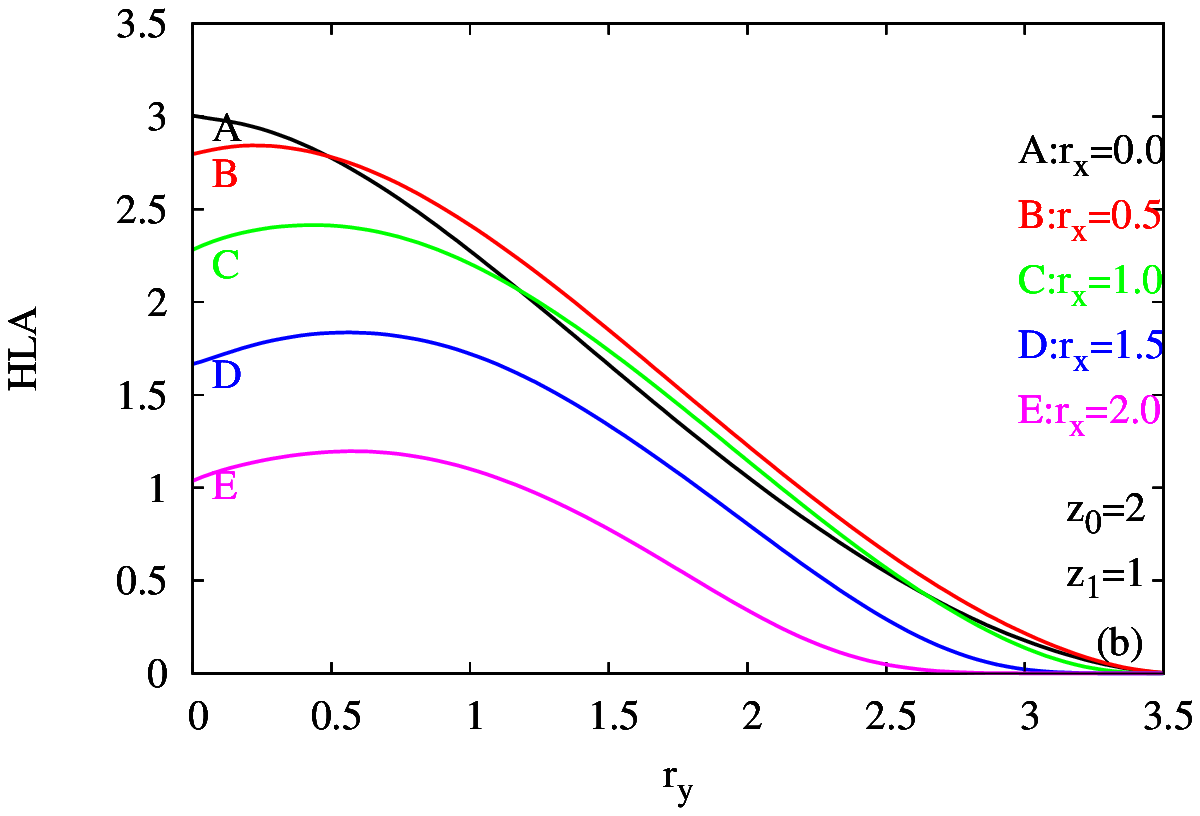, width=7.0cm}
\epsfig{file=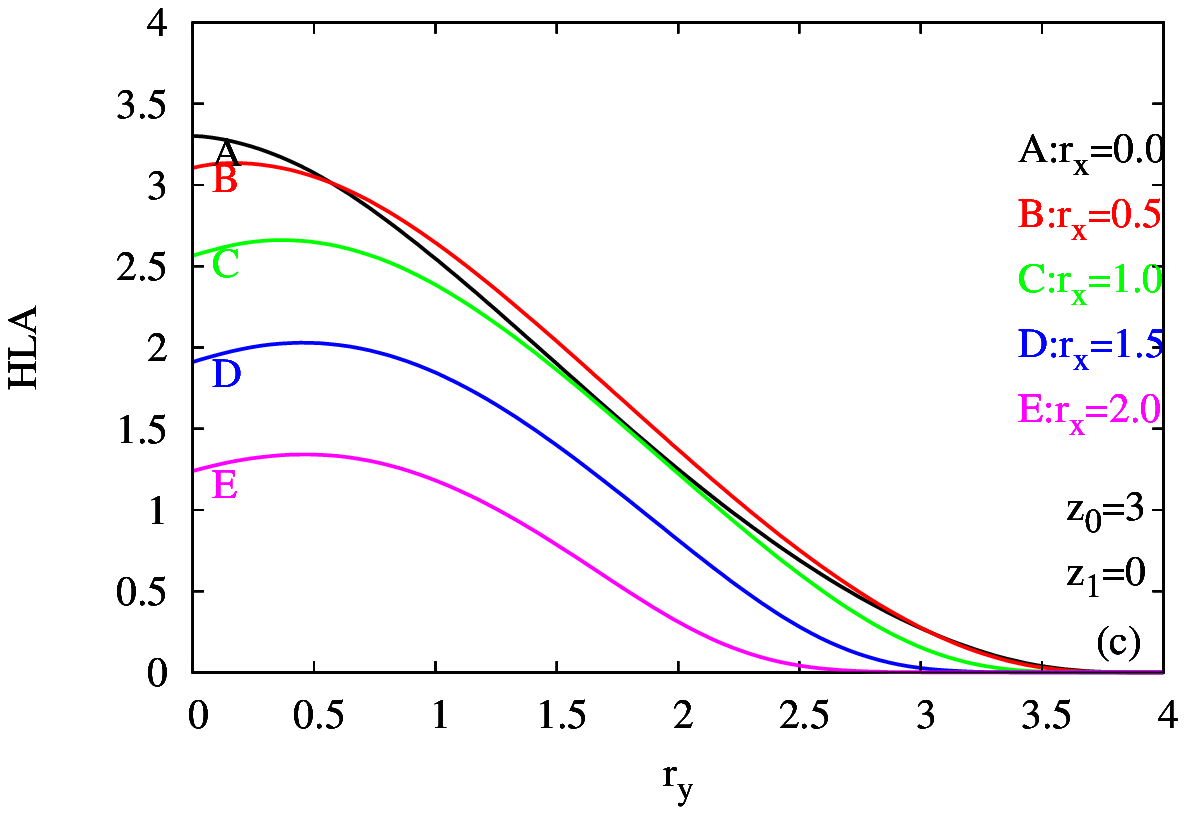, width=7.0cm}
\end{center}
\caption{The variation of the hysteresis loop areas with $r_y$ for the anisotropic quantum Heisenberg model with selected values of $r_x$ for (a) honeycomb (b) Kagome and (c) square lattices. The temperature is $T=T_c/2$} \label{sek3}\end{figure}

In Fig. \ref{sek3}, the variation of the HLA with $r_y$ for the anisotropic quantum
Heisenberg model with selected values of $r_x$ for three different lattices can be seen.
As seen in Fig. \ref{sek3}, the behavior of the HLA for the whole lattice types is qualitatively similar.  For a certain $r_x$, rising $r_y$ first slightly increases the HLA then decreases it
(except the curve for the $r_x= 0.0$ ). All HLA curves have a broad maximum in a similar $r_y$ region and reaches to zero
at a specific (which depends on the lattice type and $r_x$) value of $r_y$. This specific value is greater
for smaller $r_x$ values. This is expected, since rising anisotropy decreases the critical temperature. Also, greater valued $r_x$ curve lies below the smaller
valued $r_x$ curve (again for $r_x\ne 0.0$ ). Since HLA is different from zero for the ferromagnetic phase, all these results are expected.
Lower anisotropy means greater critical temperature, i.e. for a certain value of $r_x$, system stays in a ferromagnetic phase for a
wider range of  $r_y$. Here, an interesting result is the broad maximum behavior of the HLA with $r_y$. Rising $r_y$ decreases the critical temperature of
the system then it is expected that for a certain temperature within the ferromagnetic region, rising $r_y$ regularly decreases the HLA. But HLA curve shows a slightly rising trend then exhibit a broad maximum and finally decreases, as $r_y$ rises.  We note that  $r_x=0.0$ curve intersects the other curves. This curve is related to the $XY$ type system and shows different behavior than
the others. For $r_x=0.0$ curve, we can mention that when $r_y$ rises, HLA decreases continuously.

\begin{figure}[h]\begin{center}
\epsfig{file=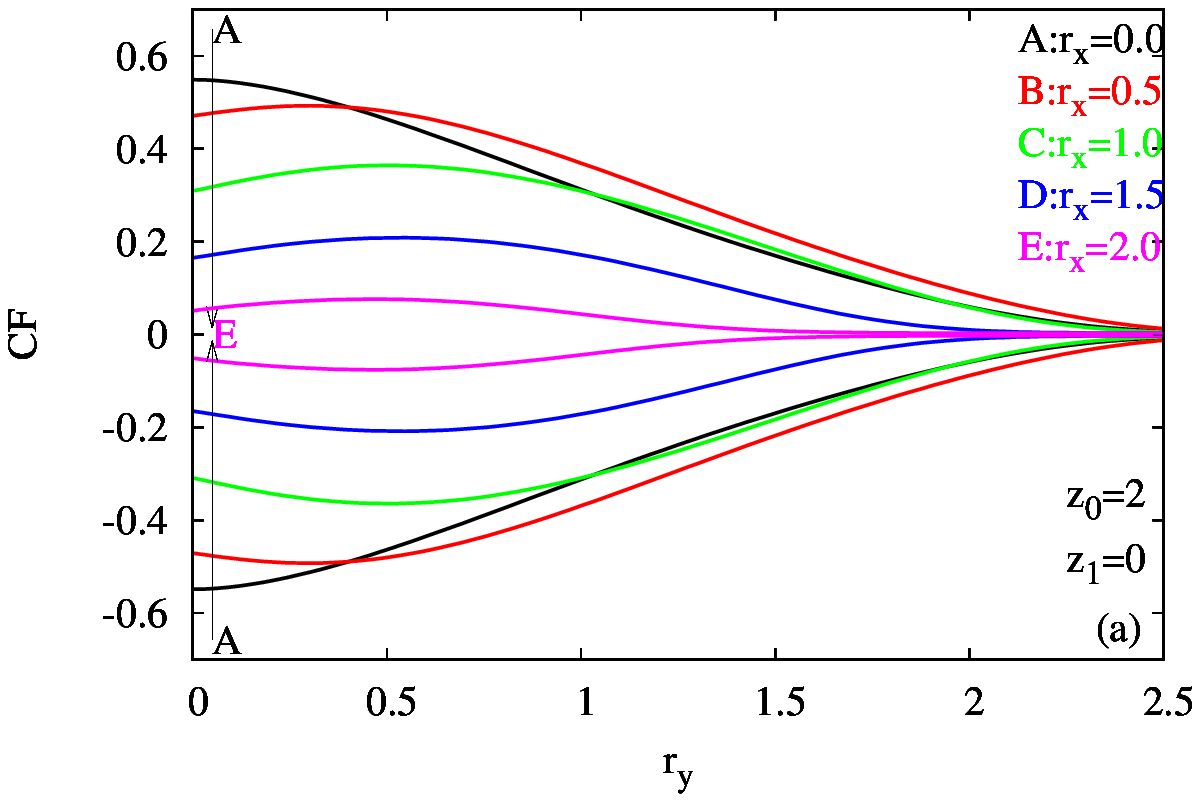, width=7.0cm}
\epsfig{file=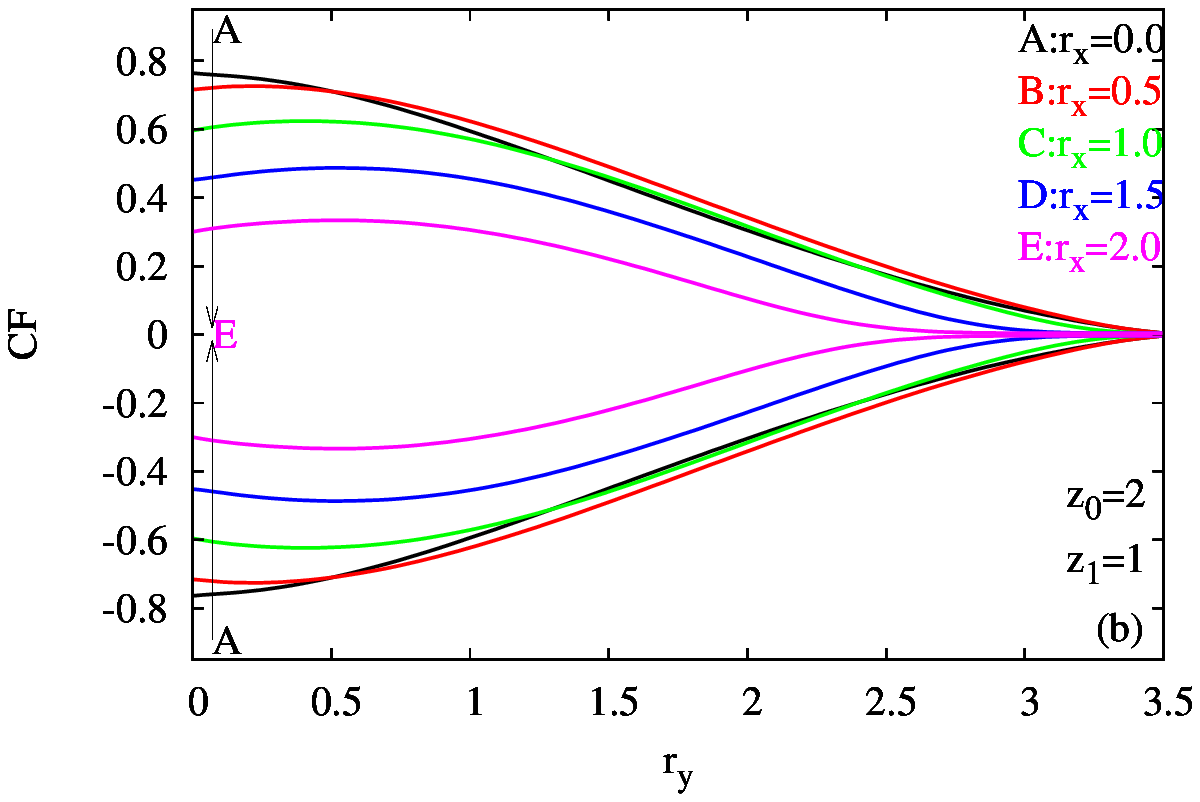, width=7.0cm}
\epsfig{file=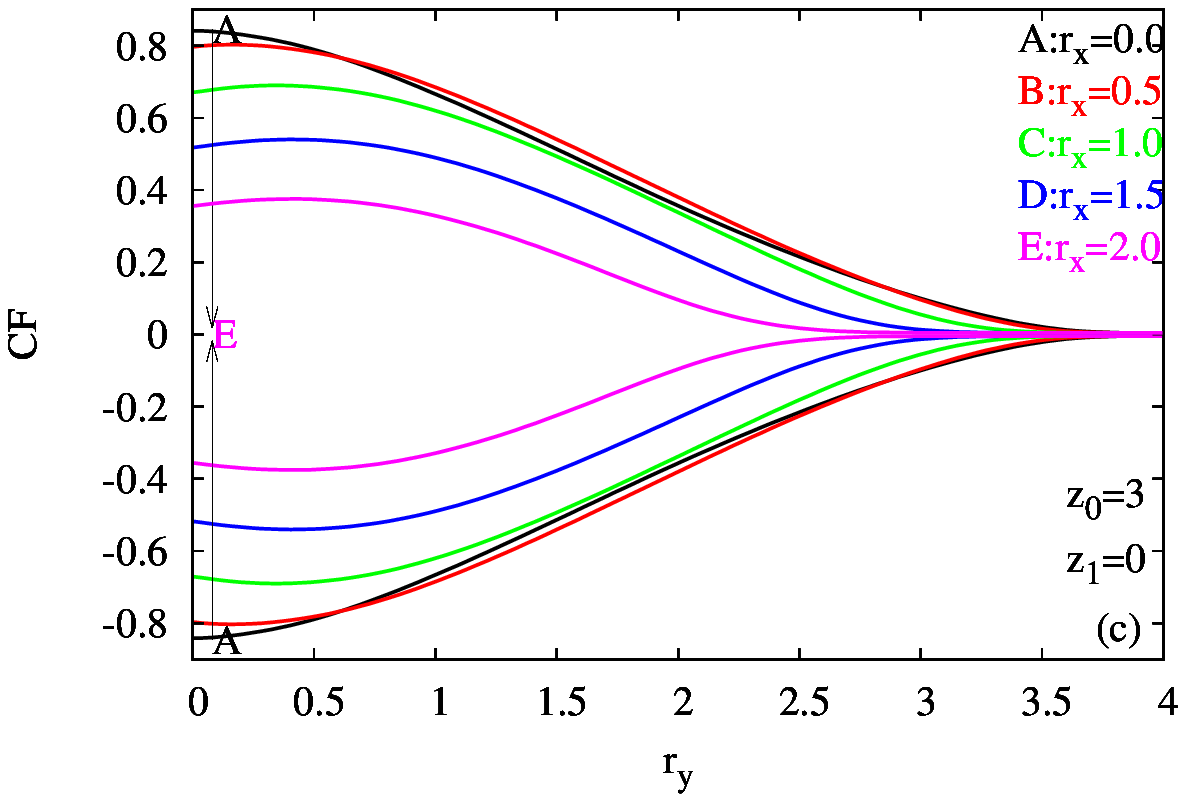, width=7.0cm}
\end{center}
\caption{The variation of the coercive field values with $r_y$ for the anisotropic quantum Heisenberg model with selected values of $r_x$ for (a) honeycomb (b) Kagome and (c) square lattices. The temperature is $T=T_c/2$} \label{sek4}\end{figure}

\begin{figure}[h]\begin{center}
\epsfig{file=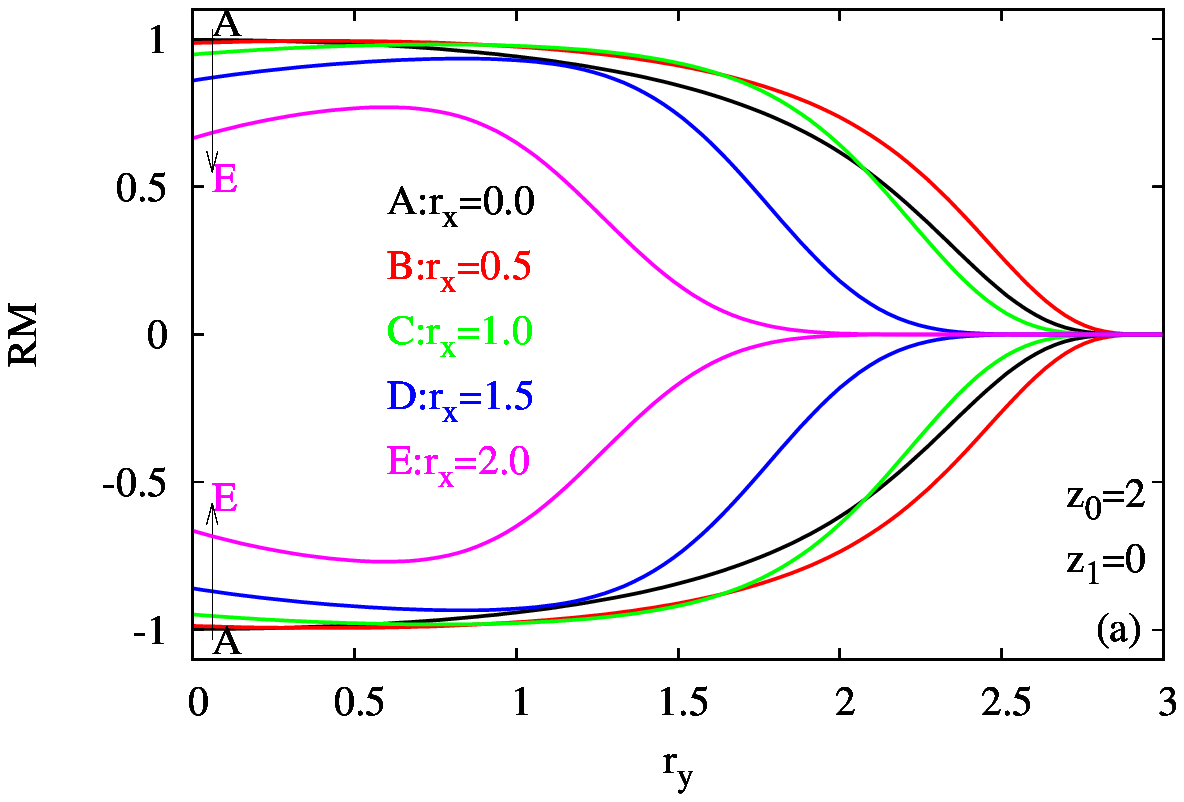, width=7.0cm}
\epsfig{file=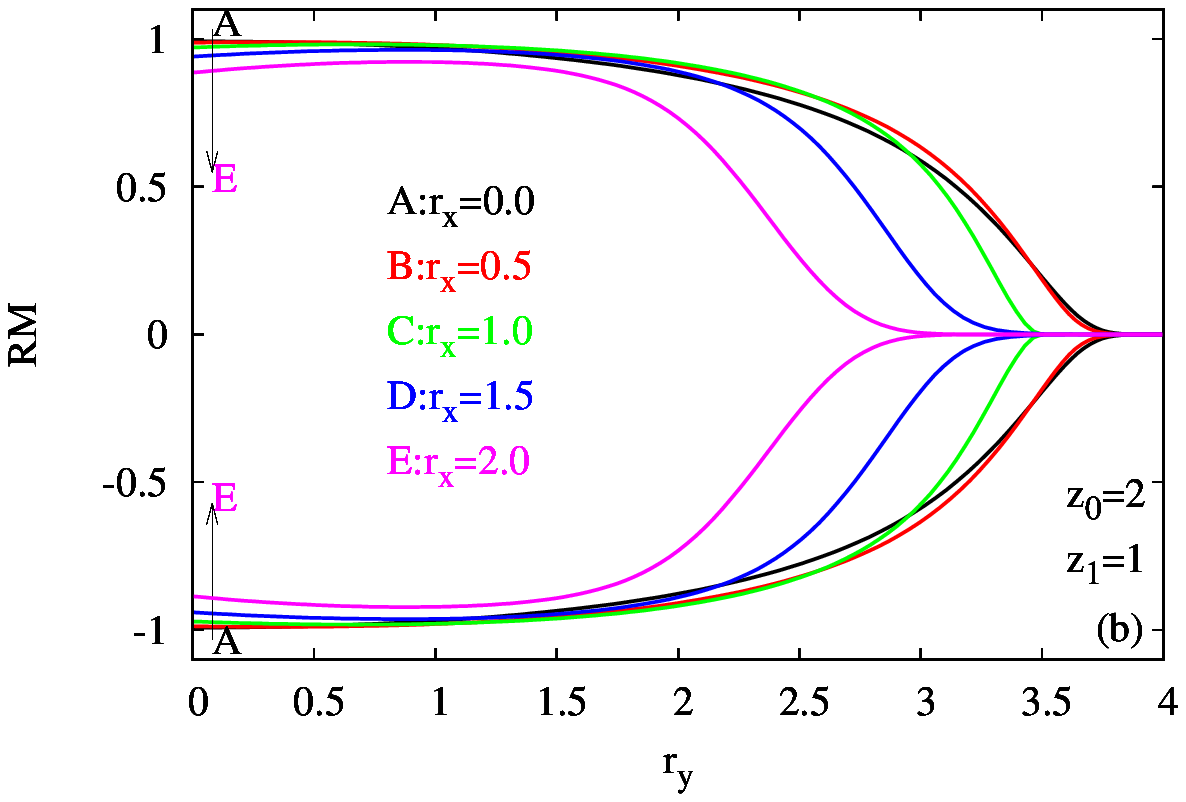, width=7.0cm}
\epsfig{file=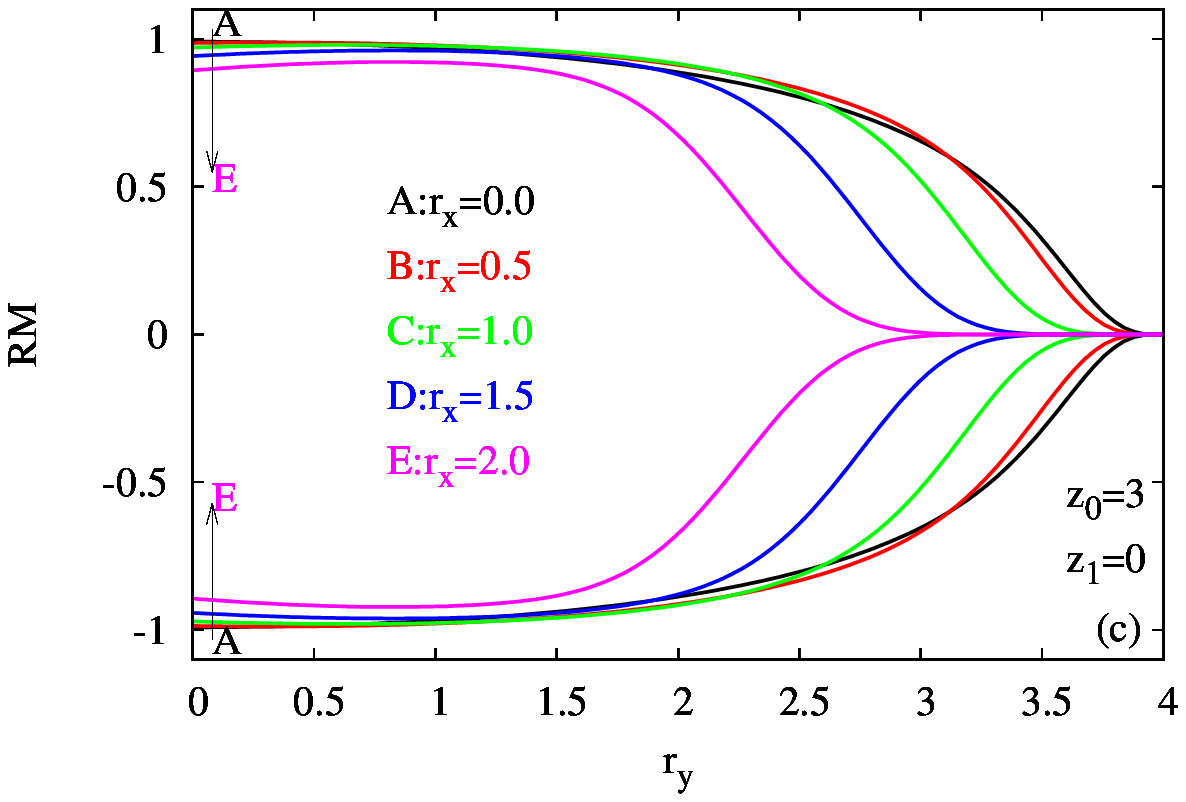, width=7.0cm}
\end{center}
\caption{The variation of the remanent magnetization values with $r_y$ for the anisotropic quantum Heisenberg model with selected values of $r_x$ for (a) honeycomb (b) Kagome and (c) square lattices. The temperature is $T=T_c/2$} \label{sek5}\end{figure}

The broad maximum behavior of the HLA curves has to show itself in the behavior of CF and RM, since HLA is related with both of them. The variation of CF and RM  with the $r_y$ for the anisotropic quantum Heisenberg model with selected values
of $r_x$ for three different lattices can be seen in Figs. \ref{sek4} and \ref{sek5},  respectively. Firstly, we can see from  Figs. \ref{sek4} and \ref{sek5} that, changing anisotropy in exchange interaction may not effect the symmetry of the hysteresis curves.
Symmetry property  of the curves in Figs. \ref{sek4} and \ref{sek5} can be regarded as a clue of the symmetry
properties of the related  hysteresis curves with respect to the origin. Actually, one can see by plotting the
hysteresis curves for different anisotropy values that all hysteresis curves are symmetric about the origin. Beside these, the
broad maximum behavior of the HLA curves with $r_y$ can be seen in these curves. We can see from the Fig. \ref{sek4} that
CF for a certain value of $r_x$ (which is different from zero) has also broad maximum behavior and from Fig. \ref{sek5} we can
observe the same behavior for the RM characteristics. The divergence rate of curves corresponding to various $r_x$ values for a honeycomb lattice is greater than the other two lattices both for CF (Fig. \ref{sek4})  and RM (Fig. \ref{sek5}). In addition, the 'sensitivity' of a certain $r_x$ value curve to the change of $r_y$ is greater for the large  $r_x$ valued curves. Again $r_x=0.0$ curve (which is related to the $XY$
type system ) intersects the other curves.

\begin{figure}[h]\begin{center}
\epsfig{file=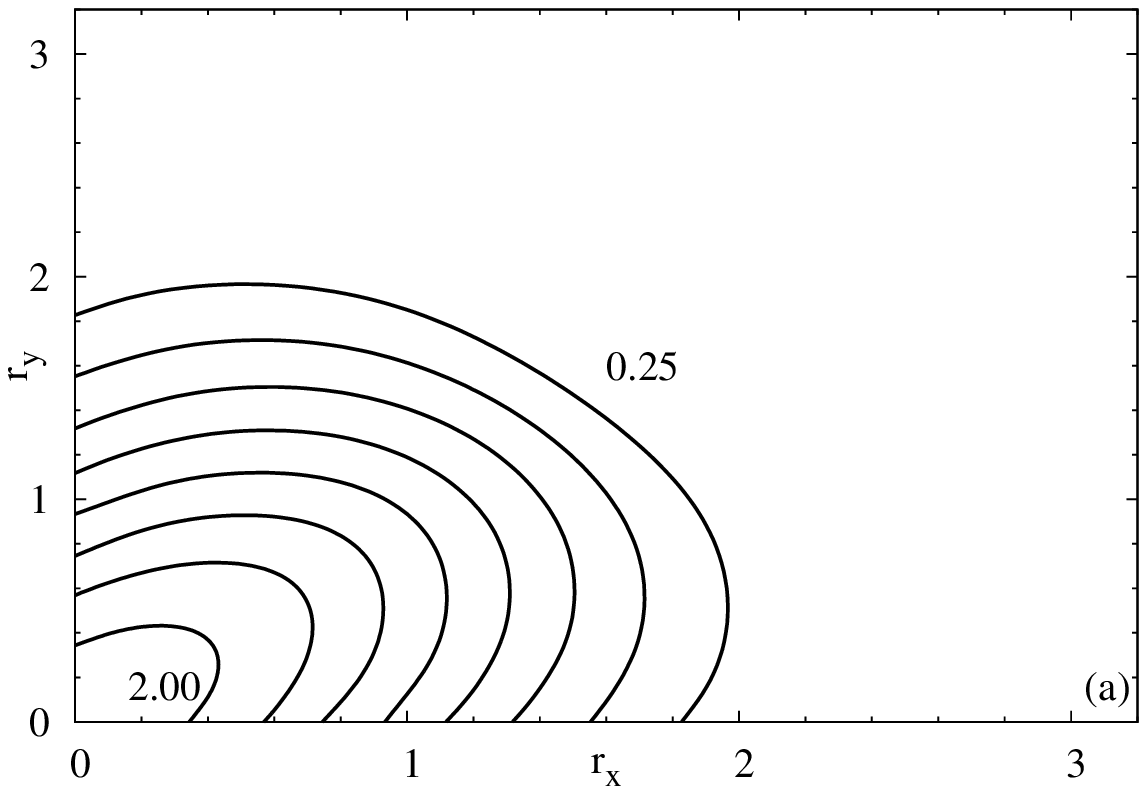, width=7.0cm}
\epsfig{file=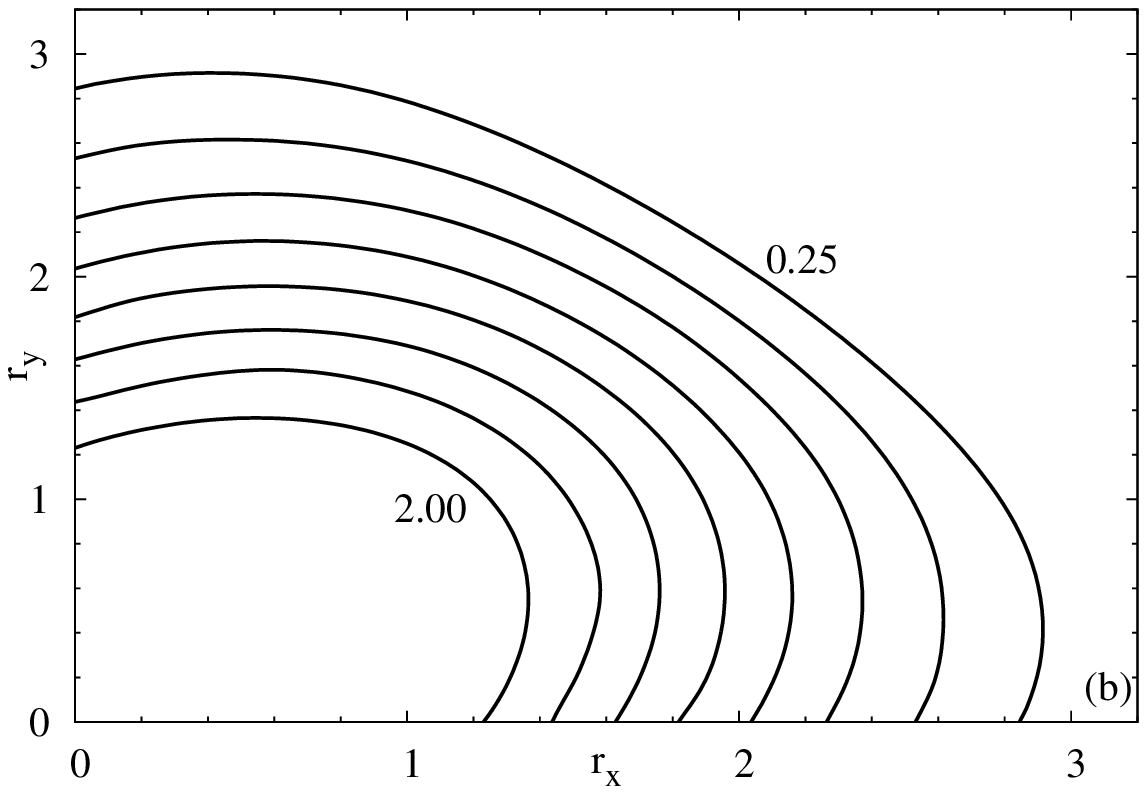, width=7.0cm}
\epsfig{file=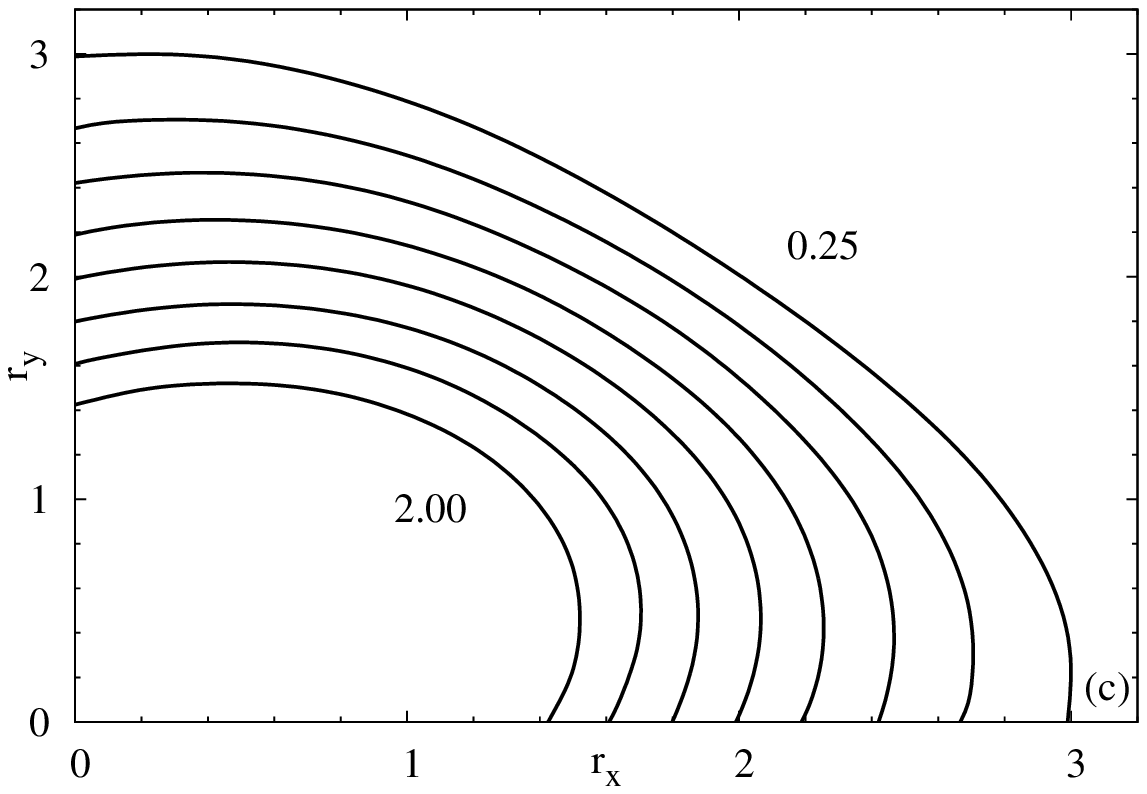, width=7.0cm}
\end{center}
\caption{Contour plots of the equally valued hysteresis loop areas in a $r_x,r_y$ plane for the anisotropic quantum Heisenberg model for (a) honeycomb (b) Kagome and (c) square lattices. The temperature is $T=T_c/2$. The inner curve stands for the value of the HLA $2.00$ while the outer one is $0.25$. Between these two curves each successive curve corresponds to an increment of $0.25$. } \label{sek6}\end{figure}

Lastly, in order to see the effect of the anisotropy in exchange interaction on the HLA from a different point of view, we draw the equally-valued HLA curves
in a $(r_x,r_y)$ plane in Fig. \ref{sek6}.

Firstly, as one can see from Fig. \ref{sek6} that, all equally valued HLA curves are symmetric with respect to $r_x=r_y$ line. This is because of the invariance of the formulation under the
transformation $r_x \leftrightarrow r_y$, as mentioned before. We can mention about three different behaviors of the $(r_x,r_y)$ pairs which holds the HLA constant, seen as the
portions of the curves   as i)  which have positive slope with respect to $r_x$ axes, ii) which have negative slope with respect to $r_x$ axes, and iii)  which have an infinite
slope with respect to $r_x$ axes. These three different behaviors correspond to three different regimes of the effect of the anisotropy in the exchange interaction:
i) both of increasing $r_x$ and $r_y$ may give the same HLA value, ii)  increasing $r_x$ and decreasing $r_y$ may give the same HLA value, iii) increasing $r_y$ and fixed $r_x$ may give the same HLA value. These three different regimes occur at some specific regions of $(r_x,r_y)$ plane depending on the lattice type. The behavior i)  occurs for the small values of $r_y$, while $r_y$ rises behavior  ii) appears instead of behavior i). Behavior iii) may occur on the square lattice for the little valued
HLA curve (see Fig. \ref{sek6} (c), $HLA=0.25$ curve). We can see from Fig. \ref{sek6}  that as we move from a greater valued HLA curve to a smaller valued HLA curve, for a certain lattice, the related curve starts to cover wider regions in the $(r_x,r_y)$ plane. This means that, greater valued HLA occurs in a narrow region of $(r_x,r_y)$ plane, and these region covers  small valued anisotropy in the exchange interaction. In addition, as we move from a gretaer valued HLA curve to a smaller valued HLA curve, behavior ii) starts to cover wider portions of the curves. This can be seen by comparing successive curves in e.g. Fig. \ref{sek6}  (b).
As we can see from Fig. \ref{sek6}, for a certain value of HLA, square lattice has longer curve
with respect to Kagome lattice and Kagome lattice has longer loop than the honeycomb lattice in the $(r_x,r_y)$ plane.


\section{Conclusion}\label{conclusion}

In conclusion, the effect of the anisotropy in the exchange interaction on the hysteresis loops within the anisotropic quantum Heisenberg model has been
investigated with the effective field theory for two spin cluster. Instead of plotting the hysteresis loops for different possible values of the
anisotropy values, they have been treated based on three fundamental properties HLA, CF and RM.

Firstly, in order to represent the behavior of the hysteresis loops with temperature, we have examined this dependence in isotropic model. The effect
of the lattice types on the behavior of the hysteresis loops with temperature have also been investigated for both ferromagnetic and paramagnetic phases.

Moreover, the effect of the anisotropy in the exchange interaction on the hysteresis loops via concerning the HLA, CF and RM features have been investigated for
three different lattice types in the ferromagnetic phase. We can say that all of these properties have qualitatively the same behavior with varying $r_y$ for selected values of reduced temperature ($T/T_c$) and $r_x$, except the case $r_x=0.0$. As $r_y$ increases then HLA (also CF and RM) slightly increases then it (also CF and RM) reaches a broad maximum region then further increasing $r_y$ causes a decrease in HLA (also CF and RM). After a certain value of the  $r_y$, these properties reaches to zero. This special $r_y$ value depends on the lattice type, reduced critical temperature and value of $r_x$. We have fixed the temperature as $T/T_c=0.5$ in our investigation for the anisotropic model.  All these properties of the hysteresis loops vanishes just after the critical temperature of the related anisotropic system (i.e. critical temperature for $r_x,r_y$ values) and we know that the critical temperature decreases with increasing anisotropy in the exchange interaction (i.e. increasing $r_x,r_y$). Hence, we conclude that for the mentioned special $r_y$ value, $(r_x,r_y)$ pair gives the critical temperature of the system as $0.5T_c$ where $T_c$ is the critical temperature of the related isotropic model. In conclusion, we can not say directly that rising anisotropy in the exchange interaction decreases the HLA, CF and RM in anisotropic quantum Heisenberg model. This is only valid for the XY model (see the curves labeled by A in Figs. \ref{sek3},\ref{sek4} and \ref{sek5}). In the case of the anisotropic quantum Heisenberg model, behavior of these properties  with $r_y$ depends on the $r_x$ value.

We hope that the results  obtained in this work may be beneficial form both theoretical and experimental point of view.

\end{document}